# Quantum Chaos[*]

Frank Steiner[**]

II. Institut für Theoretische Physik, Universität Hamburg
Luruper Chaussee 149, D-22761 Hamburg, Germany

### Abstract

A short historical overview is given on the development of our knowledge of complex dynamical systems with special emphasis on ergodicity and chaos, and on the semiclassical quantization of integrable and chaotic systems. The general trace formula is discussed as a sound mathematical basis for the semiclassical quantization of chaos. Two conjectures are presented on the basis of which it is argued that there are unique fluctuation properties in quantum mechanics which are universal and, in a well defined sense, maximally random if the corresponding classical system is strongly chaotic. These properties constitute the quantum mechanical analogue of the phenomenon of chaos in classical mechanics. Thus *quantum chaos has been found*.

---

[*] Invited contribution to the Festschrift *Universität Hamburg 1994: Schlaglichter der Forschung zum 75. Jahrestag* (Ed. R. Ansorge) published on the occasion of the 75th anniversary of the University of Hamburg, Dietrich Reimer Verlag, Hamburg 1994.

[**] Supported by Deutsche Forschungsgemeinschaft under contract No. DFG–Ste 241/4–6.

## 1. Einstein's Problem of 1917

The study of *quantum chaos* in complex systems constitutes a very fascinating and active branch of present-day physics, chemistry, and mathematics. It is not well-known, however, that this field of research was initiated by a question first posed by Einstein [1] during a talk delivered in Berlin on May 11, 1917, concerning the relation between classical and quantum mechanics of strongly chaotic systems. This seems historically almost impossible since *quantum* mechanics was not yet invented, and the phenomenon of *chaos* was hardly acknowledged by physicists in 1917.

While we are celebrating the seventy-fifth anniversary of our alma mater, the "Hamburgische Universität" [2], which was inaugurated on May 10, 1919, it is interesting to have a look upon the situation in physics in those days. Most physicists will probably characterize that time as the age of the *old* quantum theory which started with Planck in 1900 and was dominated then by Bohr's ingenious, but paradoxical model of the atom and the Bohr-Sommerfeld quantization rules for simple quantum systems. Some will associate those years with Einstein's greatest contribution, the creation of general relativity culminating in the generally covariant form of the field equations of gravitation which were found by Einstein in the year 1915 (and, independently, by the mathematician Hilbert at the same time).

In his talk [1] in May 1917, Einstein studied the Bohr-Sommerfeld quantization conditions $\int p_i \, dq_i = 2\pi n_i \hbar$, $i = 1, \ldots, l$, for systems with $l$ degrees of freedom where the $q_i$ are the coordinates and the $p_i$ their conjugate momenta. $\hbar$ denotes Planck's constant divided by $2\pi$, and the $n_i$ are integer quantum numbers. (Einstein calls the quantum conditions "Sommerfeld-Epsteinsche Quantenbedingung".) He emphasized that the products $p_i \, dq_i$ are in general not invariant and thus the quantum conditions have no invariant meaning, but rather depend on the choice of the coordinate system in which the classical motion is separable (if at all). By analyzing a simple example, the two-dimensional motion of a particle under an attractive central force, Einstein found a general coordinate-invariant formulation of the quantum conditions ($k = 1, 2, \ldots, l$)

$$I_k := \frac{1}{2\pi} \oint_{L_k} \sum_{i=1}^{l} p_i \, dq_i = n_k \hbar \;, \qquad (1)$$

noticing that the line integrals of the one-form $\sum_i p_i \, dq_i$ taken over a complete set of topologically inequivalent ("irreducible") closed loops $L_k$ are invariant. In contrast to the original version of the quantization conditions, it is not necessary to perform explicitly the separation of variables; indeed, one need not require the motion to be separable, but only to be multiply periodic. However, Einstein pointed out that conditions (1) can only be written down in the case of very special systems for which there exist $l$ integrals of the $2l$ equations of motion of the form $R_k(p_i, q_i) = $ const., where the $R_k$ are algebraic functions of the $p_i$, such that the relevant manifolds in $2l$-dimensional phase space have the shape of $l$-dimensional tori. In modern terminology, these systems are called *integrable* *systems*. (Here and in the following, we are only considering Hamiltonian systems, that is motion governed by Newton's equation without dissipation. The $l$ constants $R_k$ are assumed to be "smooth enough" and to be *in involution*, i.e., their Poisson brackets with each other vanish. See Arnold [3], and Lichtenberg and Lieberman [4] for further details.) As a result, the trajectories of integrable systems wind round these $l$-dimensional tori which in turn causes the motion of integrable systems to be very *regular* in the sense that even the long-time behaviour is well under control. Indeed, in integrable systems, trajectories with neighbouring initial conditions separate only as some power of time.

Einstein was the first physicist who realized the important rôle played by the invariant tori in phase space which he called "Trakte". He said [1]: "Man hat sich den Phasenraum jeweilen in eine Anzahl ‚Trakte' gespalten zu denken, die längs $(l-1)$ dimensionaler ‚Flächen' zusammenhängen, derart, daß in dem so entstehenden Gebilde interpretiert, die $p_i$ eindeutige und (auch beim Übergange von einem Trakt zum anderen) stetige Funktionen sind; diese geometrische Hilfskonstruktion wollen wir als ‚rationellen Phasenraum' bezeichnen. Der Quantensatz soll sich auf alle Linien beziehen, die im rationellen Koordinatenraume geschlossene sind."

However, the integrable systems forming the standard "textbook systems" with their clockwork predictability are not typical, that is "almost all" dynamical systems are *non-integrable* in the sense that there exist no constants of motion besides the energy and therefore no invariant tori in phase space. *Ergodicity* implies that almost all trajectories fill—in the absence of invariant tori—the whole $(2l-1)$-dimensional energy-surface densely. Today, our knowledge of classical dynamics is very rich [3,4], and most natural scientists begin to appreciate the importance of *chaos* in complex systems. It is now commonly recognized that generic systems execute a very irregular, *chaotic* motion which is unpredictable, that is the trajectories depend sensitively on the initial conditions such that neighbouring trajectories in phase space separate at an exponential rate.

Einstein [1] made the crucial remark that the absence of tori excludes the formulation of the quantum conditions (1) and, furthermore, that this applies precisely to the situation encountered in classical statistical mechanics where one describes the motion of colliding atoms or molecules in a gas, "denn nur in diesem Falle ist die mikrokanonische Gesamtheit der auf e i n System sich beziehenden Zeitgesamtheit äquivalent." In his "Nachtrag zur Korrektur" [1], he referred, as an example, to Poincaré in connection with the three-body problem, and he concluded: "..., und es versagt die SOMMERFELD-EPSTEINsche Quantenbedingung auch in der hier gegebenen, etwas erweiterten Form."

At the time when Einstein gave his talk, he probably was the most famous living physicist. He was a professor at the University of Berlin (with the right but not the obligation to teach!), a member of the Königlich Preußische Akademie der Wissenschaften and, shortly afterwards (October 1, 1917), he became director of the newly founded institute "für physikalische Forschung" of the Kaiser-Wilhelm-Gesellschaft zur Förderung der Wissenschaften (today the institute is called Max-Planck-Institut für Physik). It is thus a remarkable historical fact that Einstein's talk, which was published without delay by the German Physical Society [1], had no influence at all on the development of physics during the next forty years! In Bohr's famous 1918 paper [5], in which the *principle of correspondence* was exposed, one finds no reference to Einstein's talk! Einstein's "torus quantization" for integrable systems was rediscovered by the mathematician Joseph Keller only in the fifties [6]. It was Fritz Reiche—Planck's assistant at the University of Berlin from 1915 to 1918—who drew Keller's attention to Einstein's talk.

One can only speculate why Einstein's deep insight into the structure of classical phase space and his recognition of the latter's importance for quantization has been ignored for such a long time. It seems quite obvious that the main reason lies in the development of quantum mechanics a few years later, starting with Heisenberg's matrix mechanics in summer 1925, Schrödinger's wave equation in spring 1926, and Heisenberg's derivation of the uncertainty principle in spring 1927. Already in 1925, before the discovery of the Schrödinger equation, Pauli in Hamburg was able to calculate rigorously [7] the energy levels of the hydrogen atom from Heisenberg's quantum mechanics, which was considered as a great success and decisive test of the new theory. (Pauli's first position in Hamburg was "wissenschaftlicher Hilfsarbeiter in theoretischer Physik" at the Institut für Theoretische Physik; the first holder of the chair for theoretical physics was Wilhelm Lenz from October 1921 to August 1956; see ref. [2], p. 290; on January 30, 1924 Pauli received the "venia legendi für theoretische Physik" from the mathematician Erich Hecke.) Whereas Heisenberg had completely eliminated the classical orbits from his theory, Schrödinger was very much influenced by classical mechanics and the analogy between "the well-known mechanical principle due to and named after Hamilton" and the "well-known optical principle of Fermat" [8]. In his paper [9], where he discovered what was later popularized as *coherent states,* Schrödinger wanted to illustrate by the example of "Planck's linear oscillator" that it is always possible to find solutions of his "undulatory mechanics" in the form of well-localized wave packets whose center of gravity oscillates without change of shape with the period of the corresponding classical motion and thus describes the classical trajectory of a point particle like, for example, the Kepler orbit of the electron in the H-atom. (For a detailed discussion of Schrödinger's paper and the rôle it played in Heisenberg's discovery of the uncertainty principle, see [10].) On October 4, 1926, Schrödinger gave a talk in Copenhagen, where Bohr had invited him together with Heisenberg. The almost fanatic discussions between Bohr and Schrödinger have been vividly described by Heisenberg in his autobiography [11]. In the long run, Bohr and Heisenberg seemed to have won the battle, and there is no doubt that the "Copenhagen interpretation" of quantum mechanics is one of the main reasons why Einstein's torus quantization was considered out of date.

Shortly after the discovery of the Schrödinger equation, a *semiclassical* approach was devised, known as the WKB-method, named after Wentzel, Kramers, and Brioullin. In the *semiclassical limit* one studies the behaviour of quantum mechanical quantities like energy levels, wave functions, barrier penetration probabilities, decay rates, or the $S$-matrix as Planck's constant tends to zero. This limit is different from the classical limit, for which $\hbar$ is precisely equal to zero, because, in general, quantal functions are non-analytic in $\hbar$ as $\hbar$ goes to zero; examples of such a behaviour can be found in eqs. (10), (14), and (17) below. Application of the WKB-method is straightforward in the case of simple separable systems if one ignores some subtleties which typically arise if the separation is carried out in non-cartesian coordinates. (These problems can be overcome if one starts from Feynman's path integral treated in a consistent way [12].) If one tries, however, to apply the WKB-method to more complicated systems, one encounters serious difficulties which were not solved during the first decades after the discovery of the Schrödinger equation. Of course, many people did not realize these problems at all since they were content with a treatment of the simplest systems being not aware of the importance of more complex, let alone chaotic systems. This is reflected in the fact that even modern textbooks on classical or quantum mechanics usually do not mention the phenomenon of chaos. Also, the appearance of ever more powerful computers has led to a widespread belief that all problems can be solved numerically and painful analytical investigations are therefore no more worthwhile to be pursued.

As already mentioned, it was Keller [6] who discovered in the fifties that a sound mathematical derivation of the semiclassical behaviour of quantum mechanics requires a detailed knowledge of the underlying classical phase space structure. In the case of *integrable* systems, he was able to give the most general semiclassical quantization rule which turned out to be exactly Einstein's torus quantization apart from corrections arising from *Maslov indices*. Today this quantization condition for integrable systems goes under the name of *EBK-quantization*, for Einstein, Brioullin, and Keller.

Before we shall have a closer look on chaos and quantum chaos, I briefly summarize the modern version of Einstein's torus quantization. (For further details, especially on the construction of semiclassical wave functions, I refer to the recent textbook by Gutzwiller [13].) As discussed before, an *integrable* system with $l$ degrees of freedom is characterized by the existence of $l$ constants of motion in involution, where one constant of motion is the total energy $E$ which is equal to the classical Hamiltonian, $H(\mathbf{p}, \mathbf{q}) = E$. (In the following, $l$-dimensional vectors will be denoted by $\mathbf{p} = (p_1, p_2, \ldots, p_l)$.) Then each orbit of the dynamical system lies on a submanifold in phase space of dimension $2l - l = l$, which has the topology of an $l$-dimensional torus. It is now possible to make a canonical transformation from the coordinates $\mathbf{q}$ and their conjugate momenta $\mathbf{p}$ to new coordinates $(\mathbf{I}, \mathbf{w})$, called *action-angle variables*. The angles $w_k$ vary from 0 to $2\pi$ and are interpreted as new coordinates, while the *actions* $I_k$ are identified with Einstein's loop variables defined in eq. (1) and play the rôle of new conjugate momenta. If $w_k$ runs from 0 to $2\pi$, it defines a closed loop $L_k$ in the original phase space variables $(\mathbf{p}, \mathbf{q})$, where $L_k$ is the $k$th irreducible circuit of the torus. Since the integrals (1) are invariant, as noticed by Einstein, the $I_k$'s are the new constants of motion. Moreover, the new Hamiltonian $\overline{H}$ is a function of the actions $I_k$ only, $\overline{H} = \overline{H}(\mathbf{I})$. Then the *EBK-quantization condition* reads

($k = 1, 2, \ldots, l$)
$$I_k = (n_k + \beta_k/4)\hbar , \qquad (2)$$

where the $n_k \geq 0$ are integer quantum numbers, and the integers $\beta_k \geq 0$ are the Maslov indices. (The motion takes place on a so-called Lagrangian manifold, and the Maslov index—which can be understood as the number of conjugate points or the *Morse index* [14] of a trajectory—is determined by the topology of the Lagrangian manifold in phase space with respect to configuration space.) It follows from (2) that the semiclassical approximation to the quantal energy levels is explicitly given by

$$E_\mathbf{n} = \overline{H}\big((\mathbf{n} + \beta/4)\hbar\big). \qquad (3)$$

The EBK-formula (3) gives the exact leading asymptotic term as $\hbar \to 0$.
Notice that Einstein's condition (1) is only correct, if all Maslov indices vanish which is in general not the case. (E.g., for the harmonic oscillator one has $\beta = 2$.) Naturally, Einstein could not know about the Morse index theorem since "Morse theory" was not yet developed [14]. That the Maslov indices are very important for the physical properties of atomic and molecular systems, is illustrated by the example of the hydrogen molecule ion which was treated by Pauli in his Ph.D. thesis in 1919 [15]. The question was whether such a molecule as the singly ionized $H_2^+$ exists, which was not yet known at that time. Pauli calculated the ground state energy by quantizing the radial motion with respect to the axis of the molecule. For the ground state, he chose $n_r = 1$ and, of course, $\beta_r = 0$, and obtained a positive energy, and thus he concluded that $H_2^+$ is only metastable. The correct values turn out to be $n_r = 0$ and $\beta_r = 2$, and the ground state energy is negative! (For further details, see [13].)

So far we have seen how Einstein's quantum condition (1) has been rediscovered and refined, leading to the general EBK-quantization rules (2) and (3) for integrable systems. The whole theory is based on Einstein's observation that the phase space of integrable systems is foliated into $l$-dimensional tori and that each orbit moves on an invariant torus. There is, however, Einstein's second important observation, which he mentioned in his talk [1], namely that *ergodic systems* possess no invariant tori and that his quantization method can therefore not be applied! In fact, it is known [13] that the phase space of strongly chaotic systems carries two mutually transverse foliations, each with leaves of $l$ dimensions. Every trajectory is the intersection of two manifolds, one from each foliation. The distance between two neighbouring trajectories increases exponentially along the unstable manifold, and decreases exponentially along the stable manifold. Obviously, the EBK-construction based on action-angle variables is no more possible and there remained the difficult task to find a semiclassical quantization method for chaotic systems. It took another decade until Martin Gutzwiller [16] opened up the royal road towards an answer to "Einstein's question: how can classical mechanics give us any hints about the quantum-mechanical energy levels when the classical system is ergodic?" [13]. "Einstein's question"—or "Einstein's problem of 1917" as I have phrased it in the title of this section—is the starting point of our modern studies of quantum chaos in complex systems. Before I shall come to a discussion of quantum chaos, I want to recall briefly a few historical facts on the development of our ideas on ergodicity and classical chaos.

## 2. Ergodicity and Chaos

We owe the early recognition of chaotic motion in nature at the end of the nineteenth century to the physicists Maxwell, Boltzmann, and Gibbs on the one hand, and to the mathematicians Poincaré and Hadamard on the other hand. Whereas Boltzmann put forward the ergodic hypothesis in 1887 [17] leading to the Boltzmann-Gibbs model of a gas as a prototype example of thermodynamics and statistical mechanics, Poincaré was mainly concerned with the three-body problem of celestial mechanics with special attention to the earth-sun-moon system [18]. Clearly, the dynamical systems studied by Boltzmann and Poincaré are of utmost importance. It turns out, however, that they are extremely complex and even today many of their properties are not understood. For the development of ergodic theory and modern chaos theory, it was therefore very important that Hadamard [19] introduced already in 1898 a dynamical system which is simple enough to be treated mathematically, and on the other hand shows the typical behaviour of irregular motion. His system has two degrees of freedom and consists of a point particle of mass $m$ which moves freely, i.e., without external forces, on a given two-dimensional surface. (The dimension could be higher than two, but two is the smallest dimension for which chaos can occur.) Before we discuss Hadamard's example, it is worthwhile to consider another class of systems first.

Let us assume that the point particle moves on a *flat* surface and, moreover, that the motion is confined to a compact domain $\Omega \in \mathbb{R}^2$ with boundary $\partial \Omega$. Then one obtains a *planar billiard* if one imagines hard walls at the boundary $\partial \Omega$. The trajectories of the particle consist of segments of straight lines with elastic reflections at $\partial \Omega$. It turns out that the billiard dynamics depends very sensitively on the boundary $\partial \Omega$. E.g., if the boundary is a circle, an ellipse, or a square, the system is integrable while a boundary of the shape of a stadium leads to a strongly chaotic system, the well-known *stadium billiard* [20]. Although the Hamiltonian of such a planar billiard is not smooth, but rather discontinuous,

$$H(\mathbf{p},\mathbf{q}) = \begin{cases} \mathbf{p}^2/2m & \mathbf{q} \in \Omega \\ \infty & \mathbf{q} \notin \Omega \end{cases}, \qquad (4)$$

the study of the classical billiard ball problem was influential for the development of modern ergodic theory. (For an early account, see [21].) In the hands of Sinai [22] it led finally to the first proof, given in 1963, that the Boltzmann-Gibbs model of a gas is ergodic.

Instead of choosing a flat surface, Hadamard [19] considered a surface with *negative Gaussian curvature*. Having in mind our discussions of quantum chaos in the next sections, it is useful to specialize already here to surfaces with *constant negative curvature* and without boundary. One then obtains compact Riemann surfaces $\mathscr{F}$ of genus $g \geq 2$ with $\text{area}(\mathscr{F}) =: A = 4\pi(g-1)$ (Gauß-Bonnet theorem). (E.g., for $g = 2$, the particle moves freely on a frictionless surface having the topology of a double torus.) Hadamard's dynamical system is described by a smooth Hamiltonian which is given by

$$H(\mathbf{p},\mathbf{q}) = \frac{1}{2m} p_i g^{ij}(\mathbf{q}) p_j, \qquad \mathbf{q} \in \mathscr{F}, \qquad (5)$$

where $p_i = m g_{ij} dq^j/dt$ are the conjugate momenta ($t \in \mathbb{R}$ denotes time), $g^{ij}$ is the inverse of the metric $g_{ij}$ which is defined by the line element $ds^2 = g_{ij} dq^i dq^j$. (Here Einstein's summation convention is used; $i,j = 1,2$.) It follows that the classical orbits are the *geodesics* on the given surface $\mathscr{F}$. In ergodic theory [23] Hadamard's dynamical system is called the *geodesic flow on* $\mathscr{F}$.

At first sight, Hadamard's system appears to be a purely mathematical model, too abstract to be relevant for physics. However, Sinai [22] translated the problem of the Boltzmann-Gibbs gas into a study of the by-now famous *Sinai billiard*, which in turn he could relate precisely to Hadamard's model of 1898! Recently, smooth *experimental* versions of Sinai's billiard have been fabricated at semiconductor interfaces as arrays of nanometer potential wells [24] and have opened the new field of *mesoscopic systems* [25].

Hadamard's great achievement was that he could prove that all trajectories in his system are *unstable* and that neighbouring trajectories *diverge* in time at a rate $e^{\omega t}$ where $\omega = \sqrt{2E/mR^2}$ is the *Lyapunov exponent*. (Here $R$ is a length scale which is fixed by the constant negative curvature, $K = -1/R^2$.) Thus he was the first who could show that the *long-time behaviour* of a dynamical system can be very sensitive to the initial conditions and therefore *unpredictable*, even though the system is governed by a deterministic law like Newton's equations as in his model. Today this sensitivity on initial data is recognized as the most striking property of systems with *deterministic chaos*. It appears that Hadamard should be considered as the true discoverer of chaos.

Today the subject of chaos is very popular which is certainly caused to a great extent by the widespread unqualified use of the colourful word "chaos" inviting for wild speculations in many directions. It has even been claimed that the discovery of chaos theory constitutes the third great revolution in physics in the twentieth century, the first two being the invention of relativity and of quantum theory. Statements of this sort are absurd and, as we have seen, historically false. While it is true that most physicists did not take notice of chaos until recently, we have seen that Hadamard opened the doors already at the end of the last century. (One may speculate what would have happened if Hadamard had used the word "chaos" already in 1898 to illustrate his findings.) It is a mere fact that "chaos theory" was for many decades almost exclusively a domain of mathematics carrying the less fancy name of "ergodic theory".

The reader may wonder whether Hadamard's paper [19] experienced the same fate as Einstein's talk [1]. This is, however, not the case! The French physicist Pierre Duhem realized the philosophical implications of Hadamard's discovery, and in a series of articles [26], which appeared in 1904 and 1905, he described Hadamard's dynamical system with poetic inspiration. In 1906, Duhem published his articles as a separate book [27] whose German translation [28] appeared already in 1908 with a foreword by Ernst Mach. The book was translated by Friedrich Adler, a close friend of Einstein's. (When Einstein moved to Zürich in 1909, he lived with his family in the same house as Adler. It would be interesting to find out whether Adler and Einstein talked about Duhem's book and Hadamard's model, since Hadamard's paper [19] contains a lot of Riemannian geometry which became later so important in Einstein's work on general relativity. In 1916 Einstein supported his pacifist friend Adler, then in jail for having shot and killed Graf Stürgkh, the prime minister of Austria.) Duhem describes Hadamard's model, i.e., the geodesic flow on $\mathscr{F}$ as follows [28]: "Denken wir uns die Stirn eines Stieres mit den Erhöhungen, von denen die Hörner und Ohren ausgehen. Verlängern wir diese Hörner und Ohren in der Art, daß sie sich ins Unendliche ausdehnen, so haben wir eine Fläche, wie wir sie studieren wollen. Auf einer solchen Fläche können die geodätischen Linien recht verschieden aussehen. (...) Die einen winden sich unaufhörlich um das rechte Horn, die anderen um das linke oder auch um das rechte resp. linke Ohr. (...) Man kann die Genauigkeit, mit der die praktischen Angaben bestimmt sind, beliebig erhöhen, man kann den Flecken, der die Anfangslage des materiellen Punktes bildet, verkleinern, man kann das Bündel, das die Richtung der Anfangsgeschwindigkeit enthält, zusammenschnüren, man wird doch niemals die geodätische Linie, die sich ohne Unterlaß um das rechte Horn dreht, von ihren ungetreuen Kameraden befreien, die, nachdem sie sich zuerst, ebenso wie erstere, um dasselbe Horn gewunden, ins Unendliche entfernen." This is a beautiful illustration of chaos, already in 1908! Duhem refers also to Poincaré and the three-body problem and asks [28]: "Werden die Gestirne des Sonnensystems unter der Annahme, daß die Lagen und Geschwindigkeiten derselben die gleichen seien wie heute, alle weiter und unaufhörlich sich um die Sonne drehen? Wird es nicht im Gegenteil geschehen, daß eines dieser Gestirne sich von dem Schwarm seiner Gefährten trennt, um sich in der Unendlichkeit zu verlieren? Diese Frage bildet das Problem der S t a b i l i t ä t  d e s  S o n n e n s y s t e m s, das Laplace gelöst zu haben glaubte, dessen außerordentliche Schwierigkeit aber die Bemühungen der modernen Mathematiker, vor allem aber die des Herrn Poincaré dartun."

At the end of this section, I should like to describe briefly another example of a strongly chaotic system whose Hamiltonian has exactly the form (5). The system is called *Artin's billiard* after the mathematician Emil Artin who studied this model here in Hamburg in 1924 [29]. Artin's billiard is a two-dimensional non-Euclidean billiard whose billiard ball table is a noncompact Riemannian surface of constant negative Gaussian curvature $K = -1$ with the topology of a sphere containing an open end (cusp) at infinity. (An infinitely long "horn" as described by Duhem.) The surface can be realized on the Poincaré upper-half plane $\mathscr{H} = \{z = x + iy \mid y > 0\}$ endowed with the hyperbolic metric $g_{ij} = \delta_{ij}/y^2$. (The coordinates of the point particle are $q_1 = x$, $q_2 = y$, where the length scale $R$ has been put equal to one.) On $\mathscr{H}$ the modular group $\Gamma = \mathrm{PSL}(2,\mathbb{Z})$ operates via fractional linear transformations, i.e., by $\gamma = \begin{pmatrix} a & b \\ c & d \end{pmatrix} \in \Gamma$, $z \in \mathscr{H}$, $\gamma z = (az+b)/(cz+d)$. In his paper, Artin studied first the motion on the surface $\Gamma \backslash \mathscr{H}$ which can represented by the modular domain, i.e., by the fundamental region $\{|z| \geq 1 \mid -1/2 \leq x \leq 1/2\} \subset \mathscr{H}$ of the modular group with appropriate boundary identifications. This region is symmetric under reflection on the imaginary axis, and thus Artin was led to consider the desymmetrized system which can be viewed as a billiard defined on the halved domain $\mathscr{F} := \{|z| \geq 1 \mid 0 \leq x \leq 1/2\}$. $\mathscr{F}$ is a non-compact triangle of finite (hyperbolic) area $\pi/6$. Let us cite Artin [29]: "Damit haben wir aber die physikalische Realisierung. Man zeichne zunächst auf der Rotationsfläche der Traktrix ein mit dem halben Moduldreieck kongruentes Dreieck. Unser mechanisches System läßt sich dann als die kräftefreie Bewegung eines Massenpunktes in diesem Dreieck interpretieren (der Punkt sei gezwungen auf der Fläche zu bleiben), der von den Dreiecksseiten elastisch reflektiert wird."

In his paper [29], Artin introduced for the first time an important approach into the theory of dynamical systems which today is known as *symbolic dynamics*. Using an idea which goes back to Gauß, Artin was able to formulate the geodesic motion as a map in terms of continued fractions. This enabled him to show that the geodesic flow on $\mathscr{F}$ is quasi-ergodic. In fact, Artin's billiard belongs to the class of so-called *Anosov systems* which represent in the hierarchy of chaotic systems the highest level revealing the most stochastic behaviour ever possible. All Anosov systems are ergodic and possess the mixing-property [23]. These systems are called *strongly chaotic*. It can be shown quite generally, that the geodesic flows on compact symmetric Riemann spaces are Anosov systems [30].

There is no doubt that Artin's work [29] on ergodicity and Pauli's solution [7] of the hydrogen atom, which was already mentioned in sect. 1, belong to the top results in science achieved at the University of Hamburg already during the first few years.

### 3. The General Semiclassical Trace Formula

In sect. 1 I have discussed in detail Einstein's semiclassical torus quantization for *integrable* systems whose most general form is given by the EBK-quantization conditions (2) and the EBK-approximation (3) to the quantal energy levels. As pointed out already by Einstein in 1917, these semiclassical quantization rules fail completely in the case of *chaotic* systems since already the basic definition (1) is meaningless for ergodic systems. (There are no invariant tori in phase space, and thus no irreducible closed loops $L_k$ can be defined.) It took more than fifty years until Martin Gutzwiller [16] did the first step towards a semiclassical theory for *chaotic* systems. Although the original *Gutzwiller trace formula* from 1971 is plagued with serious divergencies and thus cannot be applied without ambiguities and numerical instabilities, it was recently found [31] that the formula can be improved and brought into a general form such that all series and integrals are *absolutely convergent*. It is the purpose of this section to describe the *general trace formula* [31] which forms a mathematically sound basis for the semiclassical quantization of chaos.

The general framework is Feynman's formulation of quantum mechanics [32] in terms of his "sum over histories" or *path integrals*. (For a recent account, see [33].)

In the semiclassical limit when $\hbar$ tends to zero, it is well-known that the leading contribution to the path integral comes from the classical orbits. Taking the trace of the time-evolution operator, the contributions come from those classical orbits which are *closed* in *coordinate* space. Gutzwiller [16] made the important observation that the trace of the energy-dependent Green's function (which is the Fourier transform of the time-evolution operator) is given by a formal sum over all classical orbits which are *closed* in *phase* space, i.e., all *periodic orbits*. The sum has only a formal meaning because there are infinitely many periodic orbits whose growth in number as function of the period is exponential for chaotic systems, and thus the sum is in general not even conditionally convergent for physical energies.

As an illustration of the semiclassical theory for chaotic systems, I shall consider planar billiards as introduced in sect. 2. For the quantal Hamiltonian $\widehat{H}$ we get from the classical Hamiltonian (4) $\widehat{H} = -(\hbar^2/2m)\Delta$, where $\Delta = \partial^2/\partial q_1^2 + \partial^2/\partial q_2^2$ is the Euclidean *Laplacian*. The hard walls at the billiard boundary $\partial\Omega$ are incorporated by demanding that the quantal wave functions $\psi_n(\mathbf{q})$ should vanish at $\partial\Omega$. Then the *Schrödinger equation* for the given quantum billiard is equivalent to the following *eigenvalue problem* of the *Dirichlet Laplacian*

$$-\frac{\hbar^2}{2m}\Delta\psi_n(\mathbf{q}) = E_n\psi_n(\mathbf{q}), \qquad \mathbf{q} \in \Omega \qquad (6)$$

$$\psi_n(\mathbf{q}) = 0, \qquad \mathbf{q} \in \partial\Omega \qquad (7)$$

$$\int_\Omega \psi_m(\mathbf{q})\psi_n(\mathbf{q})\, d^2q = \delta_{mn}. \qquad (8)$$

The following properties of this eigenvalue problem are standard: there exists only a discrete spectrum corresponding to an infinite number of bound states whose energy levels $\{E_n\}$ are strictly positive, $0 < E_1 \leq E_2 \leq \ldots, E_n \to \infty$. The corresponding wave functions $\psi_n(\mathbf{q})$ can be chosen real. Moreover, it follows from (6) that the eigenvalues scale in $\hbar$, $m$, and $R$ in the form $E_n = (\hbar^2/2mR^2)\epsilon_n$, where $\epsilon_n$ is dimensionless and independent of $\hbar$, $m$, and $R$. (Here $R$ denotes an arbitrary, but fixed length scale.) This implies that the semiclassical limit corresponds to the limit $E_n \to \infty$ and thus requires a study of the highly excited states, i.e., of the *high energy behaviour* of the quantum billiard. (Notice that the semiclassical limit is identical to the *macroscopic* limit, $m \to \infty$, where the mass of the atomic bouncing ball is becoming so heavy that one is dealing with a macroscopic point particle. In the following, I shall use "natural" units: $\hbar = 2m = R = 1$. Occasionally, however, $\hbar$ will be reinserted in order to identify the *perturbative* and *non-perturbative* contributions, respectively, in semiclassical expressions.)

The mathematical problem defined in eqs. (6), (7), and (8) is rather old, being the eigenvalue problem of the *Helmholtz equation* describing a *vibrating membrane* with clamped edges. Indeed, several membrane problems corresponding to the *integrable* billiard case have already been solved in the last century: the rectangular membrane by Poisson in 1829, the equilateral triangle by Lamé in 1852, and the circular membrane by Clebsch in 1862. However, the problem turns out to be highly non-trivial in cases where the classical bouncing ball problem is chaotic. In fact, no explicit formula is known for the energy levels or the wave functions in the chaotic case.

Recently, it has been realized [34] that the Schrödinger equation (6) describes according to Maxwell's equations also the TM-modes of a flat microwave resonator whose base has the shape of a billiard domain $\Omega$, if $(c/2\pi)(2mE_n/\hbar^2)^{1/2}$ is identified with the frequency $\nu_n$. (Here $c$ denotes the speed of light.) In typical experiments carried out so far, this identification holds for the lowest frequencies up to 20 GHz. These experiments can be considered as *analogue experiments* for quantum chaos, although they have nothing to do with quantum mechanics but rather deal with the *classical wave properties* of *electrodynamics*. However, since the two different physical situations are described by the same mathematics, it seems appropriate to speak about *wave chaos* in analogy to *quantum chaos* in case the base of the resonator has the shape of a chaotic billiard. The *universal signatures of quantum chaos*, to be discussed in sect. 4, can then be directly translated into corresponding *universal signatures of wave chaos*. I am quite sure that the properties of wave chaos will find important practical applications in the near future, e.g., in *electrical engineering* and *accelerator physics*.

For the following discussion I shall assume that the billiard domain $\Omega$ has been chosen in such a way that the corresponding classical system is *strongly chaotic*, i.e., ergodic, mixing, and is a so-called K-system [23]. Moreover, all periodic orbits are unstable and isolated. The periodic orbits are characterized by their *primitive length spectrum* $\{l_\gamma\}$ where $l_\gamma$ denotes the *geometrical length* of the primitive periodic orbit (p.p.o.) $\gamma$. Multiple traversals of $\gamma$ have lengths $kl_\gamma$, where $k = 1, 2, \ldots$ counts the number of repetitions of the p.p.o. $\gamma$. Let $\mathbf{M}_\gamma$ be the *monodromy matrix* of the p.p.o. $\gamma$, where $|\text{Tr}\,\mathbf{M}_\gamma| > 2$, since all orbits are (direct or inverse) *hyperbolic*. (This implies that all Lyapunov exponents $\lambda_\gamma$ are strictly positive. For details, see [13].) Moreover, let us attach to each p.p.o. $\gamma$ a *character* $\chi_\gamma \in \{\pm 1\}$ assuming that the Maslov index of $\gamma$ is even. ($\chi_\gamma$ depends on the focusing of the trajectories close to the p.p.o. $\gamma$ and on the boundary conditions on $\partial\Omega$.) Then *Gutzwiller's trace formula* [16] for the trace of the resolvent of $\widehat{H}$ (i.e., the trace of the Green's function) reads

$$\text{Tr}\,(\widehat{H} - E)^{-1} = \sum_{n=1}^\infty \frac{1}{E_n - E} \underset{(\hbar \to 0)}{\sim} \bar{g}(E) + g_{\text{osc}}(E), \qquad (9)$$

where $\bar{g}(E)$ denotes the so-called "zero length contribution" which comes from direct trajectories going from $\mathbf{q}'$ to $\mathbf{q}''$ whose length tends to zero if $\mathbf{q}'' \to \mathbf{q}'$. The contribution from the periodic orbits is given by the formal sum

$$g_{\text{osc}}(E) = \frac{i}{2\hbar\sqrt{E}} \sum_\gamma \sum_{k=1}^\infty \frac{l_\gamma \chi_\gamma^k e^{ik\sqrt{E}l_\gamma/\hbar}}{|2 - \text{Tr}\,\mathbf{M}_\gamma^k|^{1/2}}. \qquad (10)$$

(Natural units are used but keeping explicitly the $\hbar$-dependence. A small positive imaginary part has to be added to the energy $E$.)

The first problem with the trace formula (9) comes form the fact that the resolvent operator $(\widehat{H} - E)^{-1}$ is not of trace class. This follows directly from *Weyl's asymptotic formula* [35] which reads for two-dimensional planar billiards with area $A := |\Omega|$

$$\lim_{n \to \infty} \frac{E_n}{n} = \frac{4\pi}{A} \hbar^2. \tag{11}$$

Hence $E_n = \mathcal{O}(n)$ for $n \to \infty$, and the sum over $n$ in eq. (9) diverges. In order to cure this problem, one could simply consider the trace of a *regularized resolvent*, for example the trace of $[(\widehat{H} - E)^{-1} - (\widehat{H} - E')^{-1}]$ where $E'$ is an arbitrary but fixed subtraction point. The real problems with the original trace formula (9) arise, however, from the periodic-orbit sum (10). Due to the exponential increase

$$N(l) \sim \frac{e^{\tau l}}{\tau l}, \qquad l \to \infty, \tag{12}$$

of the number $N(l)$ of p.p.o. $\gamma$ whose lengths $l_\gamma$ are smaller than or equal to $l$, the infinite sum over $\gamma$ in eq. (10) is in general divergent. Since the divergence problems are a consequence of the exponential law (12) and thus of the existence of a *topological entropy* $\tau > 0$, they are not just of a formal mathematical nature but rather a direct signature of classical chaos in quantum mechanics. A positive entropy $\tau$ is the most important global property of a strongly chaotic system which expresses the fact that the information about the system is lost exponentially fast. We therefore see that the periodic-orbit expression (10) has only a formal meaning. In order to cast the semiclassical approach into a sound theory, it is necessary to replace the divergent sum (10) by a generalized periodic-orbit sum which is absolutely convergent.

Before I shall come to a discussion of the general trace formula, I should like to mention that the divergence of (10) is really a fortunate circumstance because a *non-convergence* of (10) would lead to an explicit semiclassical formula for the energy levels which in general would be completely wrong!

The first important point is to decide on the appropriate variable to work with. For the planar billiards discussed here, it turns out that the natural variable is not the energy $E$, but rather the *momentum* $p := \sqrt{E}$. In order to derive a convergent trace formula we are thus led [31] to study the analytic continuation of the (regularized) trace of the resolvent in the complex $p$-plane. It is not difficult to see that the regularized periodic-orbit sum is absolutely convergent in the upper-half $p$-plane as long as one stays beyond the so-called *entropy barrier* $\sigma_a := \tau - \widehat{\lambda}/2$, i.e., if $\mathrm{Im}\, p > \sigma_a$. Here $\widehat{\lambda} > 0$ denotes a certain asymptotic average of the Lyapunov exponents defined by

$$\frac{1}{N(l)} \sum_{l_\gamma \leq l} e^{-\lambda_\gamma l_\gamma / 2} \sim e^{-\widehat{\lambda} l / 2}, \qquad l \to \infty. \tag{13}$$

It turns out that $\sigma_a$ is strictly positive. It is then clear that the sum (10) has serious convergence problems.

In ref. [31] generalized periodic-orbit sum rules have been derived by considering the trace of a rather general function of the Hamiltonian $\widehat{H}$ instead of the trace of the resolvent, that is $\mathrm{Tr}(\widehat{H} - E)^{-1}$ has been replaced by $\mathrm{Tr}\, h(\widehat{H}^{1/2})$, where $h(p)$ is a suitable *spectral function*. We are now able to formulate the

GENERAL TRACE FORMULA [31]

for the class of planar billiards introduced before. Let $h(p)$, $p \in \mathbb{C}$, be any function which satisfies the following three conditions:
   a) $h(p)$ is an even function, $h(-p) = h(p)$;
   b) $h(p)$ is analytic in the strip $|\mathrm{Im}\, p| \leq \sigma_a + \epsilon$, $\epsilon > 0$;
   c) $h(p) = \mathcal{O}(p^{-2-\delta})$, $|p| \to \infty$, $\delta > 0$.
Then the leading asymptotic form of the trace of $h(\widehat{H}^{1/2})$ as $\hbar \to 0$ is given by

$$\sum_{n=1}^{\infty} h(p_n) \underset{(\hbar \to 0)}{\sim} -\int_0^\infty dp\, \mathcal{N}(p)\, \frac{dh(p)}{dp} + \frac{1}{\hbar} \sum_\gamma \sum_{k=1}^\infty \frac{l_\gamma \chi_\gamma^k\, g(kl_\gamma/\hbar)}{e^{k\lambda_\gamma l_\gamma/2} - \sigma_\gamma^k\, e^{-kl_\gamma l_\gamma/2}}, \tag{14}$$

where

$$g(x) := \frac{1}{2\pi} \int_{-\infty}^\infty dp\, e^{ipx} h(p) \tag{15}$$

denotes the Fourier transform of $h(p)$. $\sigma_\gamma$ is the sign of the trace of the monodromy matrix $\mathsf{M}_\gamma$, and $0 < p_1 \leq p_2 \leq \ldots$ parameterize the energy levels in the form $p_n := \sqrt{E_n}$. Moreover, $\mathcal{N}(p)$ denotes Weyl's improved asymptotic formula as will be defined below. Under the conditions a) to c), all series and the integral in eq. (14) converge *absolutely*. (Notice that a) to c) are *sufficient* conditions.)

It is obvious that the general periodic-orbit sum in eq. (14) is for a given function $h(p)$ in general non-analytic in $\hbar$ at $\hbar = 0$. Thus I call this term the *non-perturbative* contribution to the trace formula. It remains to discuss the integral term in eq. (14) that I call the *perturbative* contribution, since it is determined by the function $\mathcal{N}(p)$ which is defined as the perturbative contribution to the eigenvalue counting function $\mathcal{N}(p) := \#\{p_n \mid p_n \leq p\}$. For planar billiards we have the following Laurent expansion in $\hbar$,

$$\mathcal{N}(p) = \frac{Ap^2}{4\pi} \frac{1}{\hbar^2} - \frac{Lp}{4\pi} \frac{1}{\hbar} + C + \mathcal{O}(\hbar), \tag{16}$$

where $L = |\partial\Omega|$ denotes the total length of the boundary $\partial\Omega$, and $C$ a constant which is determined by the curvature of the boundary and by the angles at the corners if the boundary consists of a finite number of smooth segments. Since the first term in eq. (16) is equivalent to Weyl's law (11), the expansion (16) is called *Weyl's improved asymptotic formula*. The Weyl asymptotics has been discussed by many authors, in particular by Marc Kac in his famous paper [36] entitled "Can one hear the shape of a drum?". (For a review, see [37].) Inserting the expansion (16) into the integral term of eq. (14), it is clear that this term corresponds to the *perturbative* contribution to the trace formula.

The general trace formula establishes a striking *duality relation* between the quantal energy spectrum $\{E_n = p_n^2\}$ and the length spectrum $\{l_\gamma\}$ of the classical periodic orbits. Since the class of functions $h(p)$ satisfying the above conditions a) to c) is rather large, the general trace formula represents an infinite number of *periodic-orbit sum rules.* At present, these sum rules provide the only substitute, appropriate for quantum systems whose classical limit is strongly chaotic, for the EBK-quantization rules applicable to integrable systems.

As an example, let us consider the function $h(p) = e^{-p^2 t}$, $t > 0$. Then we obtain the trace of the *heat kernel* (using again natural units, but keeping $\hbar$)

$$\operatorname{Tr} e^{-\widehat{H}t/\hbar} = \operatorname{Tr} e^{(\hbar t)\Delta} = \sum_{n=1}^\infty e^{-E_n t/\hbar} \tag{17}$$
$$\underset{(\hbar \to 0)}{\sim} \frac{A}{4\pi\hbar t} - \frac{L}{8\sqrt{\pi\hbar t}} + C + \frac{1}{\sqrt{4\pi\hbar t}} \sum_\gamma \sum_{k=1}^\infty \frac{l_\gamma \chi_\gamma^k \, e^{-k^2 l_\gamma^2/4\hbar t}}{e^{k\lambda_\gamma l_\gamma /2} - \sigma_\gamma^k \, e^{-k\lambda_\gamma l_\gamma/2}}.$$

Here a few remarks are in order: i) The time $t$ and $\hbar$ enter only in the combination $\hbar t$. Thus the semiclassical limit is equivalent to the limit $t \to 0+$. Furthermore, if $\hbar t$ is replaced by $\beta$, one obtains the *partition function* of statistical mechanics, and thus the semiclassical limit corresponds to the *high-temperature limit* $\beta \to \infty$ in thermodynamics. ii) The periodic-orbit contribution in (17) vanishes exponentially as $t$ tends to zero, and one obtains the heat kernel asymptotics

$$\sum_{n=1}^\infty e^{-E_n t/\hbar} \sim \frac{A}{4\pi\hbar}\frac{1}{t} - \frac{L}{8\sqrt{\pi\hbar}}\frac{1}{\sqrt{t}} + C + \mathcal{O}(\sqrt{t}) \tag{18}$$

as $t \to 0+$. One observes that the coefficients $A$, $L$, and $C$ of eq. (18) are identical to the coefficients in Weyl's asymptotic formula (16). This does not happen by chance, but reflects a deep relation between the heat kernel asymptotics and the Weyl asymptotics. This is actually the basis of the modern approach to determine the terms in the Weyl formula. In the recent literature, the coefficients of the small-$t$ expansion of the trace of the heat kernel are sometimes called Seeley coefficients. They are topological invariants of $\Omega$, where $\Omega$ can be a general manifold. (See [36–38] for more details.) These results can also be generalized to the case of unbounded "horn-shaped" billiards [39]. If one identifies the energies $E_n$ with the frequencies $\nu_n$ of a vibrating membrane ($E_n \sim \nu_n^2$), one infers from (18) that a perfect ear can hear the area $A$ and the boundary length $L$ of the membrane that is one can hear the shape of a drum [36]. iii) It should be noticed that the asymptotic expansion (16) is identical for integrable and chaotic systems, and thus the Weyl asymptotics does not contain any information about possible fingerprints of classical chaos in quantum mechanics. iv) Notice that the Weyl formula (16) is simultaneously an asymptotic expansion in $p$ since $p$ and $\hbar$ appear only in the combination $p/\hbar$. This does not imply, however, that the error term in the counting function, $\mathcal{N}(p) - \overline{\mathcal{N}}(p)$, decreases as $\mathcal{O}(1/p)$ for $p \to \infty$. In fact, it is expected that this error term increases, but determination of its exact order is an extremely difficult problem in mathematics even for integrable systems. As an example, I refer to the error term in the famous circle problem which will be mentioned in remark iii) at the end of section 4.

The trace formula (14) represents a semiclassical approximation for planar billiards. It turns out, however, that it is *exact* for Hadamard's model that is for the geodesic flow on compact Riemann surfaces of genus $g \geq 2$, which is described by the classical Hamiltonian (5), since it is then identical with the famous *Selberg trace formula* [40]. With some modifications one also obtains an exact Selberg trace formula for Artin's billiard [41]. In analogy to the construction of Artin's billiard discussed in section 2, a compact Riemann surface can be defined as $\Gamma\backslash\mathcal{H}$, where $\Gamma$ is a discrete subgroup of $\operatorname{PSL}(2,\mathbb{R})$. On the upper-half plane $\mathcal{H}$, $\Gamma\backslash\mathcal{H}$ can be realized as a fundamental domain $\mathcal{F}$ of $\Gamma$ on $\mathcal{H}$. For $g \geq 2$, $\mathcal{F}$ can be chosen as a simply connected region whose boundaries are $4g$ geodesic segments. The classical Hamiltonian of the geodesic flow on $\Gamma\backslash\mathcal{H}$ is given by (5) with the hyperbolic metric $g_{ij} = \delta_{ij}/y^2$. Then the Schrödinger equation has the same form as in eq. (6), where $\Delta$ is now the Laplace-Beltrami operator, which on $\mathcal{H}$ simply reads $\Delta = y^2(\partial_x^2 + \partial_y^2)$. One has to impose periodic boundary conditions, $\psi(\gamma z) = \psi(z)$ for all $\gamma \in \Gamma$ and $z \in \mathcal{H}$. Then the energy spectrum is discrete, $0 = E_0 < E_1 \leq E_2 \leq \ldots$. In the case of Artin's billiard, one has the same Schrödinger equation, but with Dirichlet or Neumann boundary conditions, respectively, on $\partial\mathcal{F}$. In the first case the spectrum is discrete, whereas in the latter case the spectrum is both continuous and discrete. (For a first introduction into hyperbolic geometry and the Selberg trace formula, see [42].)

Our detailed studies of the Selberg trace formula and the semiclassical trace formula (14) carried out during the last few years have given us many insights into the complex properties of quantum systems whose classical limit is strongly chaotic. I cannot review these works here. The reader is referred to the recent literature [43–49].

### 4. Universal Signatures of Quantum Chaos

In sections 1 and 2 we have seen that the phenomenon of chaos in classical dynamics is a generic property of complex systems. The most striking property of deterministic chaos is the sensitive dependence on initial conditions such that neighbouring trajectories in phase space separate at an exponential rate. As a result, the long-time behaviour of a strongly chaotic system is unpredictable.

There arises the basic question whether this well-established phenomenon of classical chaos manifests itself in the quantum world in an analogous phenomenon which could be called "quantum chaos." By this we mean the following. Given a classical dynamical system which is strongly chaotic, is there any manifestation in the corresponding quantal system which betrays its chaotic character? The first place where one should seek for a possible chaotic behaviour in quantum mechanics seems to be the long-time behaviour in analogy to the classical case. It turns out, however, that the large-time limit in quantum mechanics is well under control due to the fundamental fact that the time-evolution operator $e^{-i\hat{H}t/\hbar}$ is unitary and thus its spectrum lies on the unit circle. Moreover, for bound state systems the spectrum is discrete and the time-evolution is therefore almost periodic in the sense of Harald Bohr's theory of almost periodic functions. This is in contrast to classical systems whose time-evolution is ruled by the Liouville operator. If the classical system is mixing and chaotic, the spectrum of the Liouvillian has a continuous part on the unit circle [23] and thus the time-evolution is unpredictable for large times. This fundamental difference between the classical and quantal time-evolution has led to the common belief that "it seems unlikely" [13] that there is anything within quantum mechanics to compare with the chaotic behaviour of classical dynamical systems. On the other hand, in the preface of his book, Gutzwiller writes [13]: "The issue is still open, and all the preliminary answers suggest that quantum mechanics is more subtle than most of us had realized."

In 1984 it has been conjectured by Bohigas et al. [50] that the statistical properties of the energy level fluctuations of chaotic systems are described by the universal laws of random-matrix theory [51]. However, it is known today that the predictions of random matrix theory agree only for short- and medium-range correlations of the quantal spectra, but fail completely for long-range correlations. This was analyzed by Michael Berry [52] using the semiclassical trace formula. Berry's semiclassical arguments suggest that one of the commonly studied spectral statistics, the so-called Dyson-Mehta [53] spectral rigidity $\Delta_3(L)$, should saturate for large $L$ in contrast to the logarithmic behaviour predicted by random matrix theory. ($\Delta_3(L)$ measures the mean-square fluctuations in the number $\mathcal{N}(p)$ of energy levels in an energy range containing on the average $L$ levels.) Moreover, it was recently found that there exists a very special class of chaotic systems, showing *arithmetical chaos*, which violate universality in energy level statistics even in the short-range regime. (See [44,45,49] for more details.) It thus appears that the properties of the spectral rigidity provide no universal signature of classical chaos in quantum mechanics.

In the following, I shall present two *conjectures*, on the basis of which I shall argue that there are unique fluctuation properties in quantum mechanics which are *universal* and, in a well-defined sense, *maximally random* if the corresponding classical system is strongly chaotic. I am convinced that these properties constitute the quantum mechanical analogue of the phenomenon of chaos in classical mechanics. Thus the claim seems to be justified that *quantum chaos has finally been found!*

The two conjectures are the following:

CONJECTURE I

(1) Let $\overline{\mathcal{N}}(p)$ be the perturbative contribution to the total number $\mathcal{N}(p)$ of energy levels $E_n = p_n^2$, $p_n \leq p$, for a typical quantum system, including all terms of the Laurent expansion in $\hbar$ up to $\mathcal{O}(\hbar^0)$ (Weyl's asymptotic formula (16) in the case of two-dimensional billiards). Then the arithmetic function $\delta_n := n - 1/2 - \overline{\mathcal{N}}(p_n) =: \mathcal{N}_{\mathrm{osc}}(p_n)$ fluctuates about zero with increasing average amplitude $a_n := a(p_n^2)$, in the sense that

$$\langle \mathcal{N}_{\mathrm{osc}}(p) \rangle := \frac{1}{\mathcal{N}(p)} \sum_{p_n \leq p} \delta_n = \mathcal{O}(p^{-1}) \qquad (19)$$

$$\langle \mathcal{N}_{\mathrm{osc}}^2(p) \rangle := \frac{1}{\mathcal{N}(p)} \sum_{p_n \leq p} \delta_n^2 = \mathcal{O}(a^2(p^2)) \qquad (20)$$

as $E = p^2 \to \infty$, where

$$a(E) = \begin{cases} E^{1/4} & \text{for integrable systems} \\ (\log E)^{1/2} & \text{for generic chaotic systems} \\ E^{1/4}(\log E)^{-1/2} & \text{for chaotic systems with arithmetical chaos.} \end{cases} \qquad (21)$$

(2) The normalized fluctuations,

$$\alpha_n := \delta_n / a_n ,$$

considered as random numbers have, as $n$ tends to infinity, a limit distribution $\mu(d\alpha)$ which is a probability distribution on $\mathbb{R}$ and is absolutely continuous with respect to Lebesgue measure with a density $f(\alpha)$, such that for every piecewise continuous bounded function $\Phi(\alpha)$ on $\mathbb{R}$, the following mean value converges

$$\lim_{p \to \infty} \frac{1}{\mathcal{N}(p)} \sum_{p_n \leq p} \Phi(\alpha_n) = \int_{-\infty}^{\infty} \Phi(\alpha) f(\alpha) \, d\alpha \qquad (22)$$

and is given by the above integral, where $f(\alpha)$ does not depend on $\Phi(\alpha)$. Moreover, the density $f(\alpha)$ satisfies

$$\int_{-\infty}^{\infty} f(\alpha) \, d\alpha = 1 , \quad \int_{-\infty}^{\infty} \alpha f(\alpha) \, d\alpha = 0 , \quad \int_{-\infty}^{\infty} \alpha^2 f(\alpha) \, d\alpha = \sigma^2 , \qquad (23)$$

where the variance $\sigma^2$ is strictly positive.

(3) For strongly *chaotic* systems, the central limit theorem is satisfied, that is the function $f(\alpha)$ is *universal* and is given by a *Gaussian* (normal distribution)

$$f(\alpha) = \frac{1}{\sqrt{2\pi} \, \sigma} e^{-\alpha^2 / 2\sigma^2} \qquad (24)$$

with mean zero and standard deviation $\sigma = 1/\sqrt{2}\pi$ or $\sigma = 1/2\pi$ in the *non-arithmetic* case corresponding to systems with time-reversal or without time-reversal invariance, respectively, and $\sigma = \sqrt{A/2\pi^2}$ in the *arithmetic* case of hyperbolic billiards with area $A$. In particular, all higher moments of the sequence $\{\alpha_n\}$ exist, where the odd moments vanish, and the even moments satisfy ($k \in \mathbb{N}_0$)

$$\lim_{p \to \infty} \frac{1}{\mathcal{N}(p)} \sum_{p_n \leq p} \alpha_n^{2k} = \frac{(2k)!}{2^k k!} \sigma^{2k}. \qquad (25)$$

(4) In contrast to the above universal situation for chaotic systems, for *integrable* systems there is in general no central limit theorem for the fluctuations, and the profile of the density $f(\alpha)$ can be very different for different systems. The higher moments of $\{\alpha_n\}$ may not converge to the moments of the limit distribution and the odd moments may not be zero such that $f(\alpha)$ is usually skew and can be both unimodal and multimodal.

CONJECTURE II

(1) Let $\psi_n(\mathbf{q})$, $n \in \mathbb{N}$, be the normalized eigenfunctions of a strongly chaotic quantum system. Then $\psi_n(\mathbf{q})$ has, as $n$ tends to infinity, a limit distribution with density $P(\psi)$, such that for every piecewise continuous bounded function $\Phi(\psi)$ on $\mathbb{R}$, the following limit converges

$$\lim_{n \to \infty} \frac{1}{A} \int_\Omega \Phi(\psi_n(\mathbf{q})) \, d^2q = \int_{-\infty}^{\infty} \Phi(\psi) P(\psi) \, d\psi \qquad (26)$$

and is given by the above integral, where $P(\psi)$ does not depend on $\Phi(\psi)$.
(2) For strongly chaotic systems, the central limit theorem is satisfied, that is the function $P(\psi)$ is *universal* and is given by a *Gaussian*

$$P(\psi) = \frac{1}{\sqrt{2\pi}\,\sigma} e^{-\psi^2/2\sigma^2} \qquad (27)$$

with mean zero and standard deviation $\sigma^2 = 1/A$, where $A$ denotes the area of $\Omega$.

A few remarks are in order: i) Conjectures I and II have been formulated for planar billiards. For hyperbolic billiards on the upper-half plane $\mathcal{H}$ one has to replace $d^2q$ in eq. (26) by $dx\,dy/y^2$. With the obvious modifications, both conjectures are hypothesized to hold for general chaotic systems. ii) Numerical tests of conjecture I have been performed for several strongly chaotic systems including arithmetical chaos: a generic hyperbolic octagon, hyperbolic triangles, and Artin's billiard for both symmetry classes. The results obtained so far strongly support conjecture I. A detailed numerical investigation will be published elsewhere. iii) That part of conjecture I which deals with the non-universal behaviour of *integrable* systems generalizes rigorous results recently obtained by Heath-Brown [54] on the famous circle problem, and by Bleher et al. [55] on the lattice point problem inside a shifted circle. iv) The conjecture that semiclassical wave functions should behave as Gaussian random functions, if the underlying classical dynamics is chaotic, has been put forward by Berry [56]. First tests with low-lying wave functions gave support to this conjecture [57]. Recently, a detailed test has been carried out [58] for highly excited quantum eigenstates and is in excellent agreement with the Gaussian behaviour (27). v) For both conjectures it is important that the central limit theorem is assumed to hold in the case of strongly chaotic systems, i.e., that the functions $f(\alpha)$ and $\Phi(\psi)$ are Gaussian distributions with mean zero and universal standard deviation. This expresses the fact of *maximal randomness* which in turn justifies it to consider these phenomena as manifestations of *quantum chaos.*

In conclusion, I believe that clear *signatures of quantum chaos* have been found which, moreover, are *universal* as expressed in the two conjectures. It remains to *derive* the conjectures from the general trace formula (14).

**Acknowledgments**

I would like to thank Ralf Aurich and Jens Bolte for helpful discussions, and Ralf Aurich, Frank Scheffler, and Gunther Steil for performing numerical tests of conjecture I. Financial support by the Deutsche Forschungsgemeinschaft under contract No. DFG–Ste 241/4–6 is gratefully acknowledged.

II. Institut für Theoretische Physik
der Universität Hamburg
Luruper Chausse 149, D-22761 Hamburg